# Reliability-Centric High-Level Synthesis

S. Tosun†, N. Mansouri†, E. Arvas†, M. Kandemir‡, and Yuan Xie‡
† {stosun,namansou,earvas}@ecs.syr.edu, Syracuse University
‡ {kandemir,yuanxie}@cse.psu.edu, Pennsylvania State University

## Abstract

*Importance of addressing soft errors in both safety critical applications and commercial consumer products is increasing, mainly due to ever shrinking geometries, higher-density circuits, and employment of power-saving techniques such as voltage scaling and component shut-down. As a result, it is becoming necessary to treat reliability as a first-class citizen in system design. In particular, reliability decisions taken early in system design can have significant benefits in terms of design quality. Motivated by this observation, this paper presents a reliability-centric high-level synthesis approach that addresses the soft error problem. The proposed approach tries to maximize reliability of the design while observing the bounds on area and performance, and makes use of our reliability characterization of hardware components such as adders and multipliers. We implemented the proposed approach, performed experiments with several designs, and compared the results with those obtained by a prior proposal.*

## 1. Introduction

With ever shrinking geometries and higher-density circuits, the issue of soft errors and reliability in system design is set to become an increasingly challenging issue for the industry as a whole. This is true for both commercial consumer applications and safety critical applications. Specifically, for high-volume low-margin consumer products, frequent soft errors can lead to expensive field maintenance. For safety critical applications, poor reliability can be catastrophic in terms of both human and equipment cost. Therefore, reliability-aware design that targets at mitigating the potential consequences of soft errors is highly desirable.

While substantial progress has been made over the years in formulating and understanding the basic concepts in high-level synthesis (HLS), most of the prior studies focused on performance-area or performance-power-area tradeoffs. In comparison, relatively fewer papers considered reliable/fault-tolerant HLS. However, considering the emerging soft error problem, it is becoming increasingly important to incorporate reliability concerns into the HLS process.

Prior work investigated soft error susceptibility of memory elements and combinational circuits [1]. It showed that combinational circuits are less susceptible to soft errors than memory elements. This is because of three major error masking effects on combinational circuits; namely logical, electrical, and latching-window masking. On the other hand, Sivakumar et al [2] demonstrate that the soft error susceptibility of combinational circuits will be comparable to that of memory circuits by the year of 2011 with the current technology trends. This significant prediction urges the computer designers for further research to reduce the soft error effects on the data-path part of their designs since the current protection techniques for combinational circuits introduce more area, power consumption, and/or performance penalty than those designed for memory elements. These observations motivate us to consider the effects of soft errors on the problem of high-level data-path synthesis and the overall reliability for the combinational part of the resulting designs. Therefore, the work proposed in this paper is orthogonal and complementary to techniques proposed for improving reliability of memory components.

In this paper, we propose and evaluate a reliability-centric HLS approach that addresses soft errors. We call our approach "reliability-centric" since it tries to maximize reliability of the resulting design while observing the bounds on area and performance. Note that this is very different from a conventional HLS framework that incorporates reliability concerns into the design in an ad-hoc fashion after the major design decisions (based on performance, area, and/or power) have already been made. This paper makes the following contributions:

◊ It describes our reliability characterization of library components such as adders and multipliers. Our library accommodates several versions of each type of resource, where each version can have different area, performance and reliability characteristics as compared to the others. In the context of this characterization, we also discuss the relationship between reliability and soft errors.

◊ It presents a reliability-centric HLS framework that operates under performance and area bounds. The framework makes use of our reliability characterization, and selects the most reliable version (implementation) for each operation (in the data-flow graph representation of the design) as long as we do not exceed the area or performance bounds.

◊ It presents an experimental evaluation of the proposed framework, and compares it to a prior study that improves reliability through redundancy. Our experimental evaluation identifies the cases where one of the techniques performs better than the other, and points out a unified approach that could merge the two techniques for increasing reliability further.

The rest of this paper is organized as follows. The next section presents a discussion of the prior work. Section 3 gives a background on soft errors. Section 4 presents the results obtained from our reliability characterization of hardware components. Section 5 explains the method used in this paper to evaluate the reliability of the overall design. Section 6 presents our approach to scheduling, resource binding, and resource allocation. Section 7 presents experimental data, showing the impact of the proposed reliability-centric approach. Section 8 concludes the paper with a summary of our contributions and summarizes our future efforts.

## 2. Related Work

Most of the prior studies on reliable design make use of component redundancy. They typically use one resource (version) for each type of operation with a fixed reliability, and the reliability is increased by adopting N Modular Redundancy (NMR). Orailoglu and Karri [3] introduced an elegant design methodology for fault-tolerant ASICs to explore the three-dimensional space of reliability, area, and performance. They presented two strategies that are based on NMR. The first strategy targets at minimizing the overall cost of the design under performance and reliability constraints, while the second one tries to maximize the reliability given the cost and performance constraints. Their technique adds an extra cost to the design that is proportional to N (in NMR) for specific resource. For example, if the design area is *S* without any redundancy and the resource with the area of *A* is duplicated (i.e., N=2), then the area of resulting design is *S+A*, excluding the area required by the result-checking circuitry and interconnects. This technique gives very good results if the cost bound permits the designer to add redundancy to the design. A similar approach is used in related studies such as [4]. In addition, these studies make use of transformations that alter the computational structure such that the original behavior is maintained. The transformation based synthesis is used to reduce the overhead introduced by redundant components.

Another method used to improve the reliability of the high-level system is to duplicate the entire structure for the self-recovering circuits. This technique is used in various studies such as [5]. After copying the entire flow graph, they used various strategies to minimize the overall area of the final design. For example, [5] exploits the freedom of operations, and schedules both the copies to reduce the area overhead. Our approach differs from these previous studies since it makes use of a reliability-characterized library that has different versions of resources with different area, performance, and reliability metrics. The library we use permits us choose the most reliable resources for a specific task. In other words, instead of increasing reliability through redundancy, we achieve reliable design by using different versions of the components (as allowed by area and performance bounds).

## 3. Background on Soft Errors

A *soft error*, also called *single event upset* (SEU), is a "glitch" in a semiconductor device [6]. These glitches are random, usually not catastrophic, and they do not normally destroy the device. Soft errors could be induced through three different radiation sources, alpha particles from the naturally occurring radioactive impurities in device materials, high-energy cosmic ray induced neutrons, and neutron induced 10B fission. Soft errors occur when the collected energy Q at a particular node is greater than a critical charge $Q_{critical}$, which results in a bit flip at that node. This concept of critical charge can be used to estimate the soft error rate (SER), as will be detailed in the next section. Figure 1 illustrates the device view and the circuit view of the bit flipping caused by a particle strike.

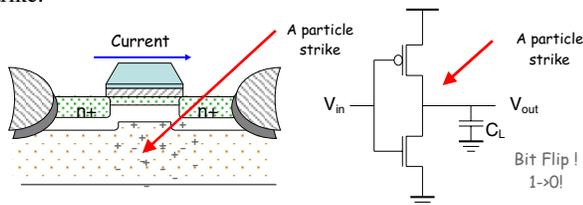

**Figure 1: Soft error phenomenon: a device view and a circuit view.**

Soft errors are the most benign form of radiation effects on the circuitry, where radiation directly or indirectly induces a localized ionization capable of upsetting internal data states. The causes for soft errors are usually outside of the designer's control. While these errors result in an upset event, the circuit itself is not damaged. Many systems can tolerate a certain degree of soft errors. For example, in a video application, soft errors can manifest themselves as missing or wrong colored bits on a display screen. These errors may or may not be noticeable or important to the user. However, when memory elements are used to control the functionality of the device, such as in an SRAM FPGA, soft errors can have a much more serious impact and lead not only to corrupt data, but also to a loss of functionality and critical failures. Soft error phenomenon in memory was known to exist as early as 1970s, and studies have been conducted to tackle this problem at the circuit level. For example, in 1996, IBM disclosed its experiments on computer electronics failure due to soft error from 1978 to 1994 [6]. Because of technology scaling, drastic shrinking in device sizes, associated with reduction in operating voltages and increase in clock frequencies, digital logic is becoming increasingly susceptible to soft errors from natural ground level radiation. Consequently, providing reliable functioning in the existence of soft errors is becoming increasingly critical.

It should be emphasized that the reliability problem is more critical for embedded systems than their general-purpose counterparts due to following reasons. First, as compared to general-purpose systems, embedded systems are generally employed in harsher environments. Second, to reduce power demands, many battery-operated embedded systems accommodate circuit/architectural-level techniques such as voltage scaling and cache shutdown, which increase the vulnerability of the entire system to soft errors. Third, the need for developing safety or mission critical embedded applications with high demands in terms of computational power under low-cost real time constraints pushed designers to explore the possibilities offered by incorporating the reliability concerns in hardware and software design of an embedded system. Therefore, reliability concern must be taken as the first-class parameter in embedded system design.

## 4. Reliability Characterization Based on Soft Errors

A key component of the proposed reliability-centric high-level synthesis effort is the library characterization for soft errors. Current state-of-the-art in library characterization [7] focuses mainly on latency, area, and power. However, it is equally important to study the soft error susceptibility of the library components so that one can conduct a tradeoff analysis between reliability and other metrics, which is critical for our purposes.

Efficient soft error fault injection and simulation techniques [8] can be used to evaluate the soft error susceptibility of a library component. For each component (such as carry-lookahead adder or carry-skip adder), each of the nodes (gates) in the netlist can be characterized individually to determine their soft error susceptibility by fault injection and simulation. After this step, by analyzing the interconnection of gates in the netlist, the overall soft error susceptibility of the design can be determined.

Our resource library has components with different area, performance, and reliability properties. The basic resources we implemented are adders and multipliers. For example, for adder implementations, we used ripple-carry adder, Brent-Kung adder, and Kogge-Stone adder, and for multiplier implementations, we used carry-save multiplier and Leap frog multiplier. In order to estimate the reliabilities of these different versions of adders and multipliers, we use a three-step approach illustrated in Figure 2.

For the first step, we derive the $Q_{critical}$ values from circuit simulation. For example, we determine the $Q_{critical}$ values for ripple-carry, Brent-Kung, and Kogge-Stone adders as 59.460e-21 C, 29.701e-21 C, and 37.291e-21 C, respectively. After finding the $Q_{critical}$ for each implementation, the soft error rate (SER) is estimated by using the expression,

$$SER \propto N_{flux} \times CS \times \exp\left\{\frac{-Q_{critical}}{Q_s}\right\},$$

proposed by Hazucha et al [9]. In this expression, $N_{flux}$ is the intensity of the neutron flux, CS is the area of cross section of the node, and $Q_s$ is the charge collection efficiency that strongly depends on doping. The other parameters, neutron flux ($N_{flux}$) and the area of cross section of the node (CS), can be chosen to be the same for different circuit implementations with the same process technology. With the assumption of uniform neutron flux and the same technology generation being used for circuit implementation, the total charge efficiency ($Q_s$) can be assumed to be the same for two circuits. Thus, the SERs for two circuits with the same technology generation can be related to each other as

$$SER_1 = SER_2 * \exp\left\{\frac{Q_{critical1} - Q_{critical2}}{Q_s}\right\}.$$

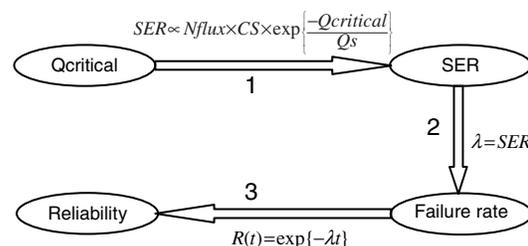

**Figure 2: Relationship between $Q_{critical}$, SER, failure rate, and reliability.**

We now need to relate the SER of each component to its reliability metric. *Reliability* is defined as the probability with which a component will perform its intended function satisfactorily for a period of time $[t_0,t]$, given that the component was working properly at time $t_0$ [10]. To calculate the reliability of a design, one needs to determine its *failure rate λ*, which is the probability with which the design will fail in the next time unit, given that it has been working properly in the current one. The reliability of a component can be related to its failure rate by the distribution function $R(t) = \exp\{-\lambda t\}$. If we assume that every soft error will result in a failure, we can use the SER of a component as its failure rate, shown as the second step in Figure 2. We can then use the reliability function to determine the reliability of a component, which is the third step in the same figure. Note that in our library characterization, the reliability of the ripple-carry adder is set to 0.999; and the reliabilities of other components are determined based on this value, using three steps depicted in Figure 2.

We laid out the circuits using the MAX layout editor tool, and used the HSPICE simulator to simulate the layouts. The normalized area and delay values for each implementation are shown in Table 1 under columns two and three, respectively. Using the steps explained above, the reliability values for these resources are estimated as shown in the fourth column of Table 1. In our experiments, we use the values given in Table 1.

| Resource type | Area (Unit) | Delay (cc) | Reliability |
|---|---|---|---|
| Adder 1 | 1 | 2 | 0.999 |
| Adder 2 | 2 | 1 | 0.969 |
| Adder 3 | 4 | 1 | 0.987 |
| Multiplier 1 | 2 | 2 | 0.999 |
| Multiplier 2 | 4 | 1 | 0.969 |

**Table 1: Area, delay, and reliability values for different adder and multiplier versions**.

## 5. Design Reliability

While the results obtained from the reliability characterization of components presented in Section 4 are important, we also need a mechanism to evaluate the reliability of an entire design built from such components. Our goal in this section is to present the model adopted in calculating the reliability of an entire design, given the reliability characterizations of individual components. Note that this is critical as it allows us to compare the two alternate designs (that implement the same functionality with different versions of resources) from a reliability perspective.

A design is typically composed of multiple components that interact with each other. The overall reliability of a design is calculated based on how these components are related to each other. Two basic reliability models are serial and parallel reliability models [10], illustrated in Figures 3(a) and 3(b), respectively.

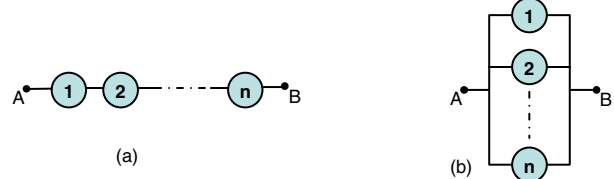

**Figure 3: Serial (a) and parallel (b) reliability models.**

In the serial model, all the components involved should succeed in order to have a system–wide (design-wide) success. As a result, the overall reliability of the system from point A to B in Figure 3(a) can be expressed as

$$R_S = \prod_{i=1}^{n} R_i .$$

In reliability engineering, the overall reliability of the parallel model between points A and B in Figure 3(b) can be found as

$$R_S = 1 - \prod_{i=1}^{n}(1 - R_i),$$

since it is assumed that only one component's success results in system-wide success. However, in the context of high-level synthesis, in order to have a successful execution of entire design, all hardware components must succeed. Consequently, to express the reliability of the design, we adopt the formula

$$R_S = \prod_{i=1}^{n} R_i$$

for parallel models as well. As an example, the reliability of the data-flow graph shown in Figure 4(a) can be expressed as $R_S = R_A * R_B * R_C * R_D * R_E * R_F$.

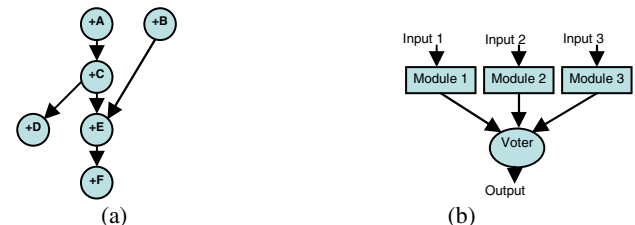

(a) (b)
**Figure 4: (a) An example data-flow graph (b) TMR structure.**

Since we want to compare our approach to a redundancy-based solution as well, let us now discuss the concept of redundancy in mathematical terms. N Modular Redundancy (NMR) [3] is a simple majority voting system that has N modules connected in parallel. TMR (Triple Modular Redundancy) is a special case of NMR illustrated in Figure 4(b). The reliability of the NMR can be expressed as

$$R_{NMR} = \sum_{i=k}^{N} \binom{N}{i} R^i * (1-R)^{N-i},$$

where $N$ is the number of components in the system and $k$ is the number of components that must succeed in order to have a successful execution. The relationship between $N$ and $k$ is given by $N = 2k - 1$. If $N \geq 3$, the structure can have the ability of fault tolerance, which is the capability of the system to continue to perform successfully after a fault occurrence [3]. If a simple duplication is used, the system can detect the fault when a fault is introduced and some recovery mechanisms such as rollback can be used to recapture the successful state of the system. In our experiments, we also used NMR structure to demonstrate the efficiency of using multiple implementations of a node (in the data-flow graph) to increase the reliability of the overall design.

## 6. Reliability-Centric Resource Allocation, Binding and Scheduling

In this section, we present resource allocation, binding and scheduling for our reliability-centric high-level synthesis approach. The problem of finding the most reliable design based on our library can be stated as follows: Given a data-flow graph $G_s(V,E)$, a resource set $R$, desired latency $L_d$, and desired area $A_d$, determine the design with the highest reliability. Note that both bounds $L_d$ and $A_d$ can prevent us from selecting the most reliable component for every operation in the data-flow graph. Each resource (version) $r$ with type $t$ in $R$ has typically a different area, performance, and reliability characteristic from the other resources (versions) of type $t$. In this section, we present our approach to determining the most reliable system by using these resources.

To illustrate the impact of using more reliable resources (instead of less reliable ones), we consider Figure 5(a) and Figure 5(b), which are two possible schedules for the data-flow graph given in Figure 4(a). For this example, we bound the latency to 5 clock cycles and the area to 4 units. For the first schedule, shown in Figure 5(a), we use only adders of type 2 (see Table 1). In contrast, for the second schedule we consider using all the adder types in Table 1. As a result, the first schedule has 4 units of total area (two adders of type 2) and a reliability of 0.82783. On the other hand, the second one has 3 units of total area (one adder of type 1 and one adder of type 2) and its reliability is 0.90713. This small example illustrates that we can have a more reliable design by using different resources with different reliability/area/performance characteristic.

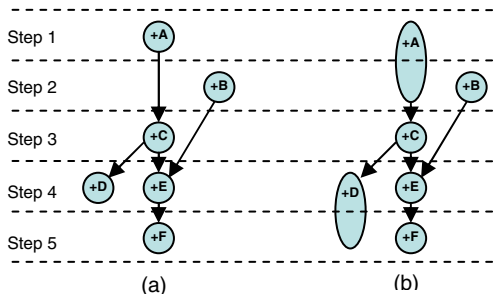

**Figure 5: Two possible schedules for the data-flow graph[*] shown in Figure 4(a).**

The overall algorithm is given in Figure 6. In this algorithm, we first find an initial solution (between lines 3 through 5), which is the most reliable one among all possible solutions. This is because this initial solution employs the most reliable version for each node in the data-flow graph. The algorithm performs resource allocation, binding, and scheduling [11] in lines 3, 4 and 5. Note that, while this algorithm can be used for both pipelined and non-pipelined data-paths, we use it here only in the context of non-pipelined circuits. The scheduling algorithm partitions the data-flow graph into the number of cycles determined by As Soon As Possible (ASAP) scheduling, and calculates the density of each partition for a specific type of operation. The total partition density is found by adding the probabilities with which a node can be scheduled within a partition. Then, it schedules an operation in the least dense partition in which the operation can be scheduled. The algorithm tries to distribute the operations evenly among the partitions so that the number of resources used in the final design is minimized. After scheduling the graph and binding the resources to each node, the algorithm returns the latency and the total area of this initial solution. As indicated earlier, this initial solution is the most reliable one since the most reliable versions are allocated for each node. However, it may not necessarily meet the latency or/and area constraints. Consequently, we may need to select a victim node and sacrifice its reliability by using a less reliable version for it. This is achieved in two steps. In the first step, given between lines 7 through 12 in Figure 6, we check if the performance constraint (bound) is met. If the latency $L$ of the initial schedule is greater than the desired latency (bound) $L_d$, then we iteratively reduce $L$ by allocating a new resource (typically a less reliable one) to a node until we reach to $L_d$. Specifically, we pick the slowest node on the critical path (Note that selecting a node which is not on the critical path will not help us reduce the initial latency value), and use a faster but potentially less reliable version for it. After this, the critical path of the current design may change. Thus, we may have to select a node from another path, which is the current critical path, in the next iteration if the current latency value is still higher than the bound ($L_d$). This process is repeated until we meet the latency bound. If all the available versions have been tried and we still could not meet the performance bound, we can conclude that it is not possible to find a solution for the graph with given latency constraint and available resources. After allocating new resources to some of the nodes, we may need to update the resource sharing if necessary (i.e., if A>A_d) to minimize the total area since we introduce new resources to the design that can increase the overall

```
1.    Find_Design (G_s(V,E),R,L_d,A_d)
2.    {
3.        Allocate the most reliable resource to each node (G,R);
4.        L=ASAP(G,R); ALAP(G,R,L);
5.        Schedule(G,R,L); Bind_Resources(G,R);
6.        A=Find_Total_Area(G,R);
7.        while(L>L_d & ∃ r' ∈ R. t_r>t_r') do
8.        {
9.            Select the node n_l on the critical path with highest delay;
10.           Allocate a resource r' to n_l such that t_r>t_r';
11.           L=ASAP(G,R); ALAP(G,R,L);Schedule (G,R,L);
12.       }
13.       Update resource sharing;
14.       A=Find_Total_Area(G,R);
15.       if(A>A_d & L_d>L)
16.       {
17.           while(L_d>L) do
18.           {   L=L+1; ALAP(G,R,L);
19.               Schedule (G,R,L); Bind_Resources(G,R);
20.           }
21.       }
22.       A=Find_Total_Area(G,R);
23.       while(A>A_d & ∃ r' ∈ R. a_r>a_r') do
24.       {
25.           Select the node n_l with the biggest area;
26.           Allocate a resource r' to n_l and to all other nodes that are
                  sharing the same resource with n_l such that a_r>a_r'
                  & t_r≥t_r';
27.           A=Find_Total_Area(G,R);
28.       }
29.       if(A>A_d or L>L_d){return no solution;}
30.       else{return total system reliability;}
31.   }
```

---
[*] In a scheduled data-flow graph, a step is a clock cycle that an operation (node) starts its execution. The type of an operation is given inside of the node with symbols such as + and * for addition and multiplication, respectively. The name (id) of the node is also given along with its type with a letter or a number.

**Figure 6: Algorithm for reliable design under performance and area constraints. $A_d$ and $L_d$ correspond to latency (performance) and area bounds, respectively.**

area of the design. This may result in sacrificing more nodes that share the same resource with the ones that we updated. After having met the latency constraint, we next calculate the total area of the design, and check whether the area bound is met. If the area $A$ of the scheduled graph is greater than the desired area $A_d$, then we make two attempts to reduce the area. We first check if there is a latency slack that could be exploited (i.e., $L<L_d$) to reduce the number of resources in the schedule. This attempt is implemented between lines 15 through 21 in Figure 6. If this is not possible, then we make our second attempt (lines 23 through 28), which is based on an idea similar to the one we employed for reducing the latency. Specifically, we select a victim node to be sacrificed based on node areas, and allocate a smaller version to it. However, when we choose a new version, we also need to check if this version increases the latency. If it does, we select an alternate node, and repeat the process until the area bound is met. If all the versions have been tried and we still could not meet the area bound, then we can conclude that it is not possible to find a solution to the given graph with given area constraint and available resources. Finally, if our algorithm returns a design that meets the area and performance bounds, we calculate the total reliability of the design using approach described in Section 5.

## 7. Experimental Evaluation

In this section, we present experimental data showing the impact of the proposed approach, and compare our results with those obtained by a prior work. In our experiments, we used several high-level synthesis benchmarks. Due to space concerns, we only give the results for three benchmark examples; namely, a 16-point symmetric FIR filter [3], a 16 point elliptic wave filter (EW) [11], and a differential equation solver (DiffEq) [12]. For the resource library, we use the values in Table 1. We first illustrate the impact of our approach on the FIR filter design. Then, we show how reliability of a design changes with respect to performance and area. Finally, we present a comparison of our approach with the solution presented in [3] on different benchmarks. We also show the results when our approach is combined with the method in [3]. Note that, except for the last experiment, our approach does not employ any redundancy.

In the first experiment, we schedule the FIR filter with two different approaches. Figure 7(a) shows the first approach that uses only one implementation for each type of operator (node). Specifically, we restrict ourselves to type 2 adders and type 2 multipliers. In comparison, the scheduling resulted from our reliability-centric approach is shown in Figure 7(b). The latency and area bounds for both the designs are 11 clock cycles and 8 units, respectively. The resulting area and reliability for the first design are 8 units (two adders of type 2 and two multipliers of type 2) and 0.48467, respectively. On the other hand, our design has a reliability of 0.78943 while the total area is the same as the first one. To reach this reliability value, our solution employs two adders of type 1, two multipliers of type 1, and one adder of type 2, resulting 8 units of total area. It must be emphasized that if we use other combinations of resources from Table 1 for the first approach, we may not be able to meet the area and/or latency bounds. For example, suppose that we used an adder of type 1 and a multiplier of type 1 to schedule the FIR filter. In this case, the minimum latency that could be achieved would be 18 clock cycles (which is larger than the 11 cycles bound). This experiment shows that having multiple versions of components with different reliability, performance, and area values can help us reach a more reliable design than an alternate scheme, which restricts itself to one type of resource only.

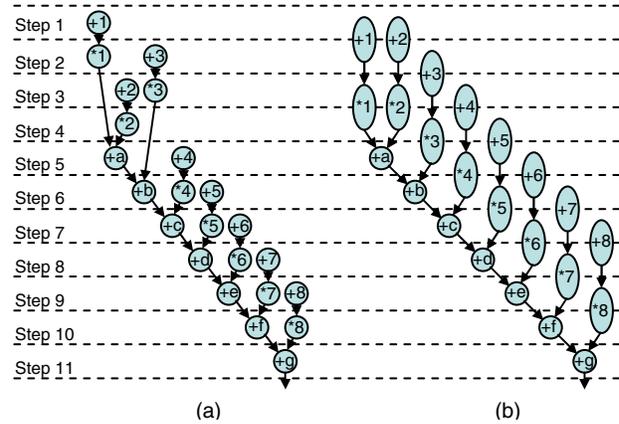

Figure 7: Two possible schedules for the FIR filter with $L_d$=11 and $A_d$=8.

In the second experiment, we demonstrate, using our approach, the tradeoff between performance and reliability and between area and reliability. We use the FIR filter in this experiment. Figures 8(a) and 8(b) plot the reliability when the latency and area values, respectively, are varied. The reliability values in Figure 8(a) are found by setting the area constraint to 8 units. As can be observed from the figure, the performance changes inversely with reliability, i.e., when we have a larger latency bound, we achieve better reliability. We study the impact of area bound on design reliability by setting a constant latency bound (10 clock cycles in this case) and varying the area bound. Figure 8(b) plots the results and shows that reliability increases proportionally with area.

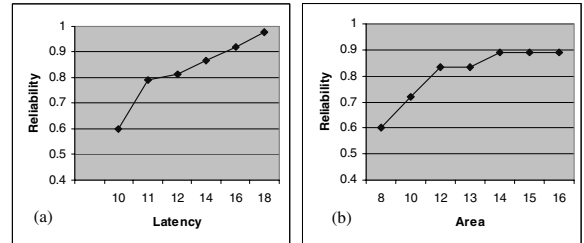

Figure 8: (a) Reliability vs performance (b) Reliability vs area.

In the final set of experiments, we compare our approach to the approach presented in [3]. We also illustrate the results obtained when these two techniques are combined. To do this, we scheduled the FIR, EW, and DiffEq benchmarks with different area and latency bounds. We started with the minimum allowable latency and area values, and then increased the area and latency bounds. For each schedule, we calculated the overall design reliability. The results are provided in Table 2(a), Table 2(b), and Table 2(c), respectively. We show the area and latency bounds used in the first two columns of these tables. In column three, we give the reliability values obtained using the technique presented in [3], and column four shows the reliability values obtained by our approach explained in this paper. Column five gives the percentage (reliability) improvements brought by our approach over the technique in [3]. Note that a negative value means that our approach generates worse result than [3] under those particular parameters. As can be seen from this column, our approach generates more reliable designs when the latency/area bounds are tighter. However, when we start to increase the area bound while keeping the latency bound constant, [3] improves the overall design reliability. With larger area bounds, it finds better results. For example, in Table 2(a), when the latency bound is 10 clock cycles, and the area bound is 9 units, [3] finds a design with the reliability of 0.48467. In comparison, with the same bounds, our

### Table 2(a): FIR filter

| Bounds | | Ref [3] | Our approach | % Imprv | Our approach +Ref [3] | % Imprv |
|---|---|---|---|---|---|---|
| $L_d$ | $A_d$ | | | | | |
| 10 | 9 | 0.48467 | 0.59998 | 23.79 | 0.59998 | 23.79 |
| 10 | 11 | 0.61856 | 0.69516 | 12.38 | 0.76572 | 23.79 |
| 10 | 13 | 0.76572 | 0.69516 | -9.22 | 0.77187 | 0.80 |
| 11 | 9 | 0.48467 | 0.78943 | 62.88 | 0.79497 | 64.02 |
| 11 | 11 | 0.61856 | 0.89798 | 45.17 | 0.98411 | 59.10 |
| 11 | 13 | 0.76572 | 0.89798 | 17.27 | 0.99102 | 29.42 |
| 12 | 9 | 0.61856 | 0.81387 | 31.58 | 0.81959 | 32.50 |
| 12 | 11 | 0.76572 | 0.90890 | 18.70 | 0.98411 | 28.52 |
| 12 | 13 | 0.78943 | 0.90890 | 15.13 | 0.99301 | 25.79 |

(a)

### Table 2(b): EW filter

| Bounds | | Ref [3] | Our approach | % Imprv | Our approach +Ref [3] | % Imprv |
|---|---|---|---|---|---|---|
| $L_d$ | $A_d$ | | | | | |
| 13 | 7 | 0.45509 | 0.70260 | 54.39 | 0.81225 | 78.48 |
| 13 | 9 | 0.67645 | 0.78463 | 15.99 | 0.97530 | 44.18 |
| 13 | 11 | 0.89005 | 0.78463 | -11.84 | 0.98805 | 11.01 |
| 14 | 7 | 0.45509 | 0.71114 | 56.26 | 0.83739 | 84.01 |
| 14 | 9 | 0.69739 | 0.79417 | 13.88 | 0.97530 | 39.85 |
| 14 | 11 | 0.94641 | 0.79417 | -16.09 | 0.98805 | 4.40 |
| 15 | 5 | 0.45509 | 0.69739 | 53.24 | 0.69739 | 53.24 |
| 15 | 7 | 0.71899 | 0.80383 | 11.80 | 0.81225 | 12.97 |
| 15 | 9 | 0.97530 | 0.80383 | -17.58 | 0.97530 | 0.00 |

(b)

### Table 2(c): DiffEq

| Bounds | | Ref [3] | Our approach | % Imprv | Our approach +Ref [3] | % Imprv |
|---|---|---|---|---|---|---|
| $L_d$ | $A_d$ | | | | | |
| 5 | 11 | 0.70723 | 0.77497 | 9.58 | 0.77497 | 9.58 |
| 5 | 13 | 0.82370 | 0.80403 | -2.39 | 0.82370 | 0.00 |
| 5 | 15 | 0.82783 | 0.80645 | -2.58 | 0.84920 | 2.58 |
| 6 | 11 | 0.70723 | 0.82370 | 16.47 | 0.82700 | 16.94 |
| 6 | 13 | 0.82370 | 0.82370 | 0.00 | 0.82783 | 0.50 |
| 6 | 15 | 0.82783 | 0.90260 | 9.03 | 0.90712 | 9.58 |
| 7 | 7 | 0.70723 | 0.90260 | 27.62 | 0.90260 | 27.62 |
| 7 | 9 | 0.82370 | 0.93054 | 12.97 | 0.93054 | 12.97 |
| 7 | 11 | 0.82783 | 0.95935 | 15.89 | 0.95935 | 15.89 |

(c)

**Table 2: Reliability values and improvements under different latency and area bounds. (a) FIR filter, (b) EW filter, (c) DiffEq.**

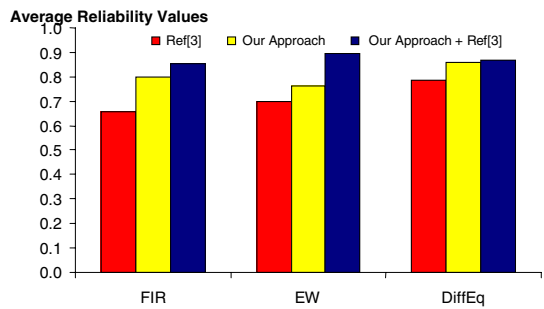

**Figure 9: Average reliability values for [3], our approach, and the combined approach for three different HLS benchmarks.**

approach obtains a reliability value of 0.59998, a 23.79% improvement over [3]. However, when the area bound is loosened to 13 under the same latency bound, [3] increases the reliability of the design to 0.76572, which means a 9.22% improvement over our approach. When we combine these two methods (i.e., when we also employ redundancy following our approach), we can have even more reliable designs under very tight bounds as shown in column six of Table 2. For the combined approach, we introduce redundancy as suggested in [3]. Note that, when we add redundancy for an operator, we use the same version selected by our reliability-centric approach as duplicate(s). For example, if we use adder of type 2 in the schedule, we also use the same type of adder as the redundant copy. The improvements brought by this combined approach over [3] are given in column seven of Table 2. This combined approach obtains a better reliability than [3], as can be seen from these tables.

In Figure 9, we give the average reliability values (over all our experiments) obtained by [3], our approach, and the combined approach for FIR, EW, and DiffEq benchmarks. Our approach brings 21.92%, 9.67%, and 9.21% overall design reliability improvements over [3] for FIR, EW, and DiffEq designs, respectively. The combined approach obtains even more reliable designs with design reliability increase of 30.33%, 28.57%, and 10.26% over [3] for FIR, EW, and DiffEq, respectively. Note that, the reliability values for these designs are found for tight area and latency values, which is desirable for system design.

## 8. Conclusions and Future Work

This paper focuses on high-level synthesis and presents a reliability-centric approach to address the growing soft error problem. The main idea behind this approach is to increase the reliability of the design as much as possible, bounded only by allowable area and latency. As opposed to the prior work on the topic, the proposed framework accommodates different versions of the same type of resource, each differing in performance, area, and/or reliability. Our experimental evaluation identifies the cases where one can expect the proposed approach to be better than the prior proposal. We also discuss how our approach can be combined with the prior work to achieve even further improvements on reliability of the design under consideration. As our future plan, we would like to compare this approach to other possible alternate schemes such as optimizing area under reliability and performance constraints, or optimizing performance under reliability and area constraints.